\documentclass[final,5p,times,twocolumn,number]{elsarticle}

\usepackage{amssymb}
\usepackage{graphicx}
\usepackage{txfonts}
\usepackage{color}
\usepackage{hyperref}
\usepackage{hypernat}
\usepackage{float}
\usepackage{bm}

\journal{Chemical Physics Letters}
\begin{document}
\definecolor{green4}{RGB}{0,139,0}

\begin{frontmatter}

\title{\textit{Ab initio} potential curves for the X~$^2\Sigma_u^+$, A~$^2\Pi_u$ and B~$^2\Sigma_g^+$ states of Ca$_{2}^+$} 

\author[]{Sandipan Banerjee\corref{cor1}}
\ead{banerjee@phys.uconn.edu}

\author[]{John A. Montgomery, Jr.}

\author[]{Jason N. Byrd}

\author[]{H. Harvey Michels}

\author[]{Robin C\^ot\'e}

\cortext[cor1]{Principal corresponding author; Fax: +1 860 486 3346}

\address{Department of Physics - University of Connecticut, Storrs, CT 06269-3046, USA.}

\begin{abstract}
We report \textit{ab initio} calculations of the X~$^2 \Sigma_{u}^+$,
A~$^2\Pi_u$ and B~$^2 \Sigma_{g}^+$ states of the Ca$_{2}^+$ dimer.
All electron CAS+MRCI calculations are performed for the X~$^2 \Sigma_{u}^+$ and B~$^2 \Sigma_{g}^+$ states,
while valence CAS+MRCI calculations using an effective core potential are used to
describe the A~$^2\Pi_u$ state.
A double well is found in the B~$^2 \Sigma_{g}^+$ state. Spectroscopic
constants, vibrational levels, transition moments and radiative lifetimes
are calculated for the most abundant isotope of calcium ($^{40}$Ca). The static dipole and quadrupole polarizabilities, and the
leading order van der Waals coefficients are also calculated for all three states.\\
\end{abstract} 


\end{frontmatter}

\section*{Introduction}

The presence of near degeneracies in the constituent atoms of diatomic
molecules can lead to a rich structure in the resulting interaction potentials.
In our recent work on the
Be$_2^+$ dimer \cite{Banerjee2010}, we showed that the nearly degenerate 2s-2p
state of beryllium leads to a complex set of low lying molecular curves,
including a double minima in the lowest $^2\Sigma_g^+$ state.  
In this Letter, we present new computational results on
the Ca$_2^+$ dimer, and demonstrate that this system also exibits
a rich manifold of low-lying molecular states. 
Both Be and Ca represent group 2A elements in the periodic table, with their valence
electronic structures (2s)$^2$ and (4s)$^2$, respectively. Thus, some similarities 
between Be$_2^+$ and Ca$_2^+$ may be expected.
The calculations presented here should help guide experimental efforts 
on cold molecular ions.\\

The last few years have seen significant interest in ultracold atom-ion
scattering \cite{Zhang2011, zhang2009} in the atomic, molecular and optical
physics community. The experimental realization of Bose-Einstein condensation
(BEC) has led to numerous applications involving charged atomic and molecular
species. The cooling and trapping \cite{Weiner1999} of charged gases at
sub-kelvin (ultracold) temperatures is a topic of growing interest. The
phenomena of charge transport like resonant charge transfer
\cite{Robin_Dalgarno2000} and charge mobility \cite{Robin2000} at ultracold
temperatures have also been studied in detail. Other emerging fields of interest
include ultracold plasmas \cite{Plasma1}, ultracold Rydberg gases
\cite{Rydberg1} and systems involving ions in a BEC \cite{Robin2002, BEC1}. \\

In this work, we begin by describing the methods used in our calculations, and follow 
with a discussion of the results, which include the potential curves of the 
X~$^2\Sigma_u^+$, B~$^2\Sigma_g^+$ and
A~$^2\Pi_u$ states and their spectroscopic constants.  We also calculate
electric dipole transition moments for the X~$^2\Sigma_u^+$ $\leftrightarrow$
B~$^2\Sigma_g^+$ and the B~$^2\Sigma_g^+$ $\leftrightarrow$ A~$^2\Pi_u$
transitions. Bound vibrational levels are computed for all the states along with
Franck-Condon overlaps and radiative lifetimes for the most abundant calcium isotope ($^{40}$Ca 96.94\%). 
We conclude with an analysis of
long range behavior, calculation of static atomic dipole and quadrupole
polarizabilities and determination of the van der Waals dispersion coefficient
C$_6$. 

\section*{Methods}

\begin{figure}
\centering
\includegraphics[clip, width=1.0\linewidth]{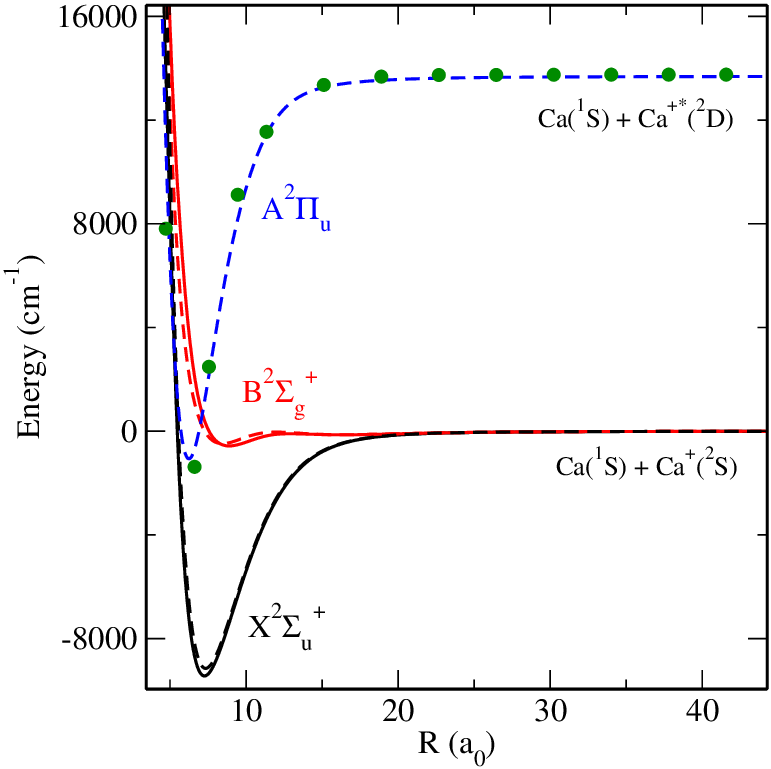}
\caption{[COLOR ONLINE] \textit{Ab initio} X~$^2\Sigma_{u}^+$ (in black), B~$^2\Sigma_{g}^+$ (in red) and A~$^2\Pi_{u}$ (in blue) states of Ca$_{2}^+$. Dashed lines show calculation with a pseudopotential, while solid lines show the results of an all electron correlated calculation. For the A~$^2\Pi_{u}$ state, results of all electron MRCI calculations are shown in green \color{green4}$\bullet$\color{black}. Note that a$_0$ is the Bohr radius (atomic unit of length).} 
\label{fig:curves2}
\end{figure}

We can express the total energy of the Ca$_2^+$ dimer at any interatomic separation R as

\begin{equation}
\label{eqn:e-tot}
E_{total} = E_{valence} + \Delta E_{core-valence} + \Delta E_{scalar-relativistic} \, .
\end{equation}

For calculation of the ground X~$^2\Sigma_{u}^+ $ and B~$^2\Sigma_{g}^+ $ states in Ca$_2^+$, the valence contribution to the total energy is calculated by a multi-reference configuration interaction (MRCI) method using a 18 orbital complete active space (CAS) wavefunction as a reference. 
The active space was chosen to include molecular counterparts of nearly degenerate 4s, 4p and 3d orbitals of Ca. The state-averaged CAS includes all doublet states correlated to Ca$^+$($^2$D) $+$ Ca($^1$S) and Ca$^+$($^2$S) $+$ Ca($^1$S) asymptotes with equal weights.
We have used the augmented correlation consistent polarized valence quintuple zeta (aug-cc-pV5Z) basis set of Peterson \cite{Peterson2002, Peterson1}. 
In order to assess the quality of MRCI, we do a comparison with a full CI calculation with aug-cc-pVTZ basis, and find out that the difference in total energy at the equilibrium separation of $7.3$ bohrs for the X~$^2\Sigma_u^+$ state of Ca$_2^+$ is $4.5$ microhartrees. At large separation ($1000$ bohrs), this difference further reduces to $1.5$ microhartrees.\\ 

The second term in Eq. (\ref{eqn:e-tot}), the correction from the core-valence contribution is estimated by,

\begin{equation}
\label{eqn:e-cv}
\Delta E_{core-valence} = [E_{RIV} - E_{Val}]_{R} -  [E_{RIV} - E_{Val}]_{R_\infty} \, .
\end{equation}

The core-valence correction $\Delta E_{core-valence}$ is the difference of
energies from a valence only (Ar core) and a restricted inner valence (RIV, Ne
core) CCSDT calculation. For this purpose we have used the correlation
consistent polarized weighted core-valence triple zeta (cc-pwCVTZ) basis set of
Koput and Peterson \cite{Peterson2002}.
To assess the convergence with basis set of the calculated core-valence
contribution,
we performed single point calculations at the equilibrium bond separation for
the ground state with a larger basis set (cc-pwCVQZ). The effect of increasing
the basis set from TZ to QZ changed the core-valence energy by $\sim$ $8$
cm$^{-1}$ at the $R_e$ of the B~$^2\Sigma_g^+$ state. These results indicate
that the core-valence contribution to the total energy is adequately converged
with the TZ basis sets.\\

The last correction term  $\Delta E_{scalar-relativistic}$ is the contribution
from relativistic effects, which for a heavy atom like Ca is significant. This
can be expressed as,

\begin{equation}
\label{eqn:e-dk}
\Delta E_{scalar-relativistic} = [E_{rel} - E_{non-rel}]_{R} -  [E_{rel} - E_{non-rel}]_{R_\infty} \, .
\end{equation}

We have used the Douglas-Kroll version of the cc-pwCVTZ basis set from Kirk
Peterson (cc-pwCVTZ-DK), and performed CCSDT calculations to estimate this
correction. The magnitude of scalar relativistic correction at the equilibrium
bond distance for the B~$^2\Sigma_g^+$ state of the Ca$_2^+$ is $\sim$ 180
cm$^{-1}$.  For Ca$_2^+$, the valence electron space contains only 3 electrons, thus the
valence CCSDT is equivalent to full CI.\\

The calculation of the A$^2\Pi_u$ state, correlating to the Ca$^+$ $3d$ atomic
level, is complicated by the near degeneracy with the Ca $4s4p$ atomic level. 
The second excited $^2\Pi_u$ state comes from an atomic asymptote of Ca $4s4p$ and Ca$^+$ $4s$,
which lies $\sim$ $1500$ cm$^{-1}$ above the A$^2\Pi_u$ asymptote.
We find, however, that valence CAS+MRCI calculations incorrectly predict the
Ca $4s4p$ and Ca$^+$ $4s$ asymptote to lie below the Ca $4s^2$ and Ca$^+$ $3d$ asymptote.
The correct ordering of the atomic energy levels is obtained when core-valence
correlation including double excitations of the inner valence electrons
are included in the correlation treatment using the cc-pwCVQZ (or better) basis set.
We expect that a balanced description of valence and core-valence interactions
in the A$^2\Pi_u$ state
would be obtained from a CAS(19,26)+MRCI calculation that includes molecular orbitals
arising from the atomic $3s$, $3p$, $3d$, $4s$ and $4p$ orbitals.
This is a much more demanding calculation than those required for the 
X~$^2\Sigma_u^+$ and B~$^2\Sigma_g^+$ states that correlate to ground states atoms.
It was found that a smaller CAS(19,21)+MRCI+Q/cc-pwCVQZ (MRCI plus
Davidson correction) calculation correlating the
$3s3p$ inner valence electrons with a $4s3d+4p_x$ valence reference was
sufficient to obtain the correct ordering of the first and second $^2\Pi_u$
molecular states.  Extending this calculation to include the entire $4p$
reference space using the cc-pwCV5Z was attempted but was too computationally
demanding for our available resources.
An alternative approach to the
multi-reference all electron calculation is to replace the argon core of the Ca
atoms with an effecive-core potential (ECP) where the effects of core-valence
correlation are including using a core polarization potential
(CPP)\cite{Fuentealba1985}.  This method was used
with great success by Czuchaj {\it et al} for Ca$_2$ ground and excited states
\cite{Czuchaj2003}.  We have performed comparisons between the all electron
calculations for the first two $^2\Sigma$ states as described above and the
valence $4s4p3d$ space MRCI using the ECP+CPP and basis set of Czuchaj {\it et al}
\cite{Czuchaj2003}.  The agreement was found to be satisfactory for the case of
the lowest $\Sigma$ states as seen in Fig. \ref{fig:curves2}.  Additionally we
have compared the A$^2\Pi_u$ state calculated using the same ECP+CPP method to
the core-valence MRCI+Q/cc-pwCVQZ calculation using the $3s3p4s3d+4p_x$ space discussed
above.  These two calculations agree very well, as demonstrated by Fig. \ref{fig:curves2}.
Because of the good agreement with the all electron
calculations and the computational limitations in performing multi-reference
core-valence correlation calculation, we have used the ECP+CPP method to
calculate the A$^2\Pi_u$ state in this paper.
We note in passing that the use of the CPP is essential; without it one
does not obtain the correct ordering of the excited atomic asymptotes.\\

All the potential curves are also corrected for basis-set superposition error (BSSE) using the standard counterpoise technique of Boys and Bernardi \cite{CPcorrection}. The BSSE was negligible ($\sim$ 2 -- 4 cm$^{-1}$) at the potential minima for the different curves. 
The MRCI valence calculations were done using the \texttt{MOLPRO 2010.1} electronic structure program \cite{MOLPRO_brief}. The core-valence CCSDT calculations were carried out using \texttt{CFOUR} (coupled-cluster techniques for computational chemistry) program \cite{CFOUR}. The scalar relativistic corrections were done at the CCSDT level of theory using the \texttt{MRCC} (multi-reference coupled cluster) program \cite{MRCC} of M. K\'{a}llay. All of the programs were running on a Linux workstation. 
All calculations employed restricted open-shell (ROHF) reference wavefunctions. Le Roy's \texttt{LEVEL} program \cite{LEVEL} has been used to calculate the bound vibrational levels, Franck-Condon factors and radiative lifetimes, discussed in the following section.

\section*{Results and Discussions}
\subsection*{Potential Curves and Spectroscopic Constants}

\begin{figure}
\centering
\includegraphics[clip, width=1.0\linewidth]{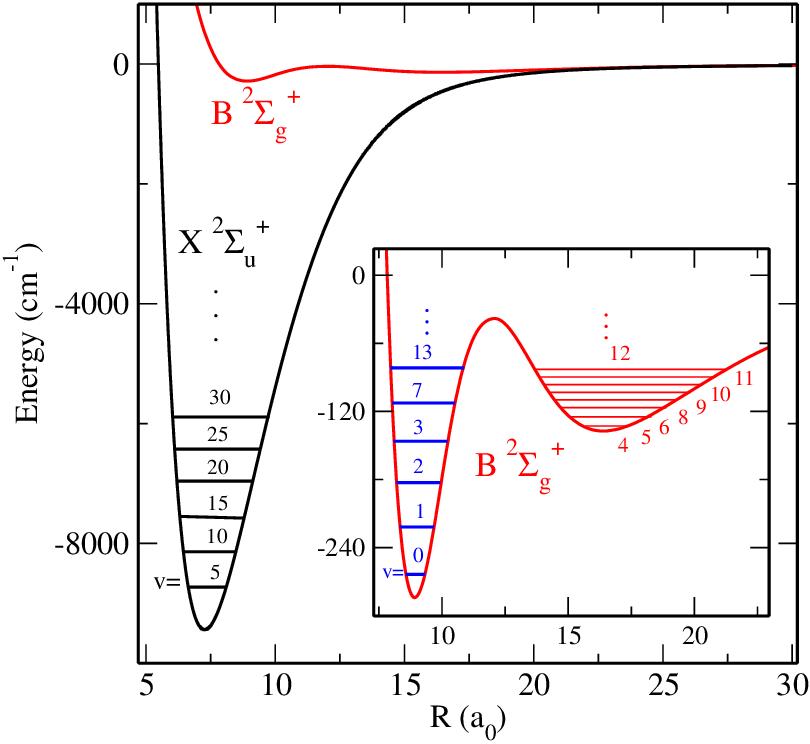}
\caption{[COLOR ONLINE] Calculated \textit{ab initio} potential curves of Ca$_{2}^+$. The inset is a magnification of the double-well nature in the B~$^2\Sigma_{g}^+$ state (in red); lowest vibrational levels in the inner well are shown in blue thick lines and outer well in red thin lines, for $^{40}$Ca.}
\label{fig:curves}
\end{figure}

\begin{table}
\centering
\caption{Calculated spectroscopic constants of Ca$_{2}^+$}
\label{tab:spec}
\scalebox{0.8}{%
\begin{tabular}{r r r r r r}\\
\hline \hline \\
State & r$_{e}$  ({\AA})  &  B$_{e}$ (cm$^{-1}) $ & $\omega_{e}$ (cm$^{-1})$ & $\omega_{e} x_{e}$ (cm$^{-1})$ & D$_{e}$ (cm$^{-1})$  \\ \\
\hline \hline \\
X~$^2 \Sigma_{u}^+$  & 3.844 & 0.056 & 127.829 & 0.071 & 9440 \\ 
Previous \cite{Czuchaj2003} & 3.773 & & 132.300 & & 9817 \\
Previous \cite{Liu1978} & 3.995 & 0.053 & 119.000 & & 8388 \\ \hline \\
B~$^2 \Sigma_{g}^+$ (Inner)  & 4.719 & 0.037 & 41.593 & 0.561 & 284 \\ 
B~$^2 \Sigma_{g}^+$ (Outer) & 8.665 & 0.011 & 8.549 & 3.839 & 137 \\  \hline \\
A~$^2 \Pi_{u}$  & 3.303 & 0.077 & 194.195 & 0.370 & 14746 \\ 
\hline \hline
\end{tabular}}
\end{table}

Fig. \ref{fig:curves} shows the \textit{ab initio} potential curves for the X~$^2\Sigma_{u}^+ $ and B~$^2\Sigma_{g}^+ $ states of Ca$_{2}^+ $ dimer. The calculated potential energy curves are corrected for the effects of basis set superposition error by the counterpoise method of Boys and Bernardi\cite{CPcorrection}. Fig.\ref{fig:curves2} shows the A~$^2\Pi_{u}$ state. We have used a standard Dunham analysis \cite{Dunham1932} to calculate the spectroscopic constants (Table \ref{tab:spec}). We calculate bound vibrational levels for the X~$^2\Sigma_{u}^+ $, B~$^2\Sigma_{g}^+ $ and A~$^2\Pi_{u}$ state for the 
$^{40}$Ca$_2^+$ dimer.\\

Unfortunately there are no experimental spectroscopic data for the ground or
excited states of the Ca$_2^+$ dimer. There are, however, some previous
theoretical studies of Ca$_2$ \cite{Czuchaj2003, Bussery2006} and Ca$_2^+$
\cite{Sullivan2011, Czuchaj2003, Liu1978}, and a comparison to the results for
the X~$^2 \Sigma_{u}^+$ state is listed in Table \ref{tab:spec}. No
spectroscopic constants have been reported for the most recent calculation by
Sullivan \textit{et al.} \cite{Sullivan2011}. Another approach for calculating interaction energies in alkaline earth
elements, is using symmetry-adapted perturbation theory (SAPT), which has been
demonstrated earlier by Patkowski {\it et al}\cite{Patkowski2007}.
The B~$^2\Sigma_{g}^+$ state has a
double well similar to that found in our recent Be$_2^+$ calculations
\cite{Banerjee2010}. Both of these wells support bound vibrational states. This
double-well nature of the B~$^2\Sigma_{g}^+$ state is most likely caused by
perturbations from an excited $^2\Sigma_{g}^+$ state. \\

\begin{figure}
\centering
\includegraphics[clip, width=1.0\linewidth]{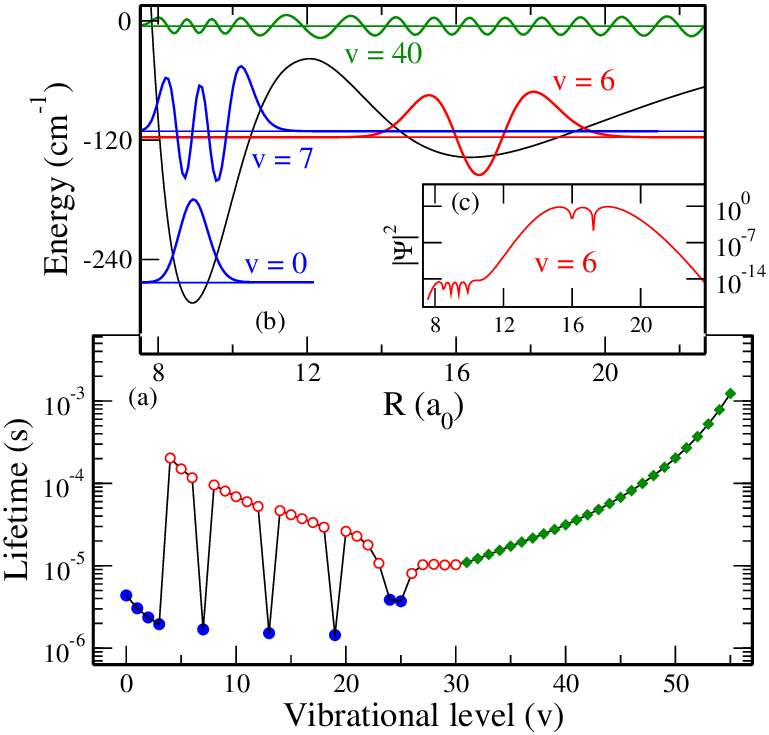}
\caption{[COLOR ONLINE] (a) Calculated radiative lifetimes of bound levels of $^{40}$Ca$_2^+$ in the B~$^2\Sigma_{g}^+$ state, on a log-scale. The shorter lifetimes (blue \color{blue}$\bullet$\color{black}) correspond to bound levels localized in the inner well, the longer lifetimes (red \color{red}$\circ$\color{black}) to levels localized in outer well, and the increasingly longer lifetimes (green \color{green4}$\Diamondblack$\color{black}) to levels spread over both wells. (b) show examples of each cases: $v$=0 and 7 in the inner well, $v$=6 in the outer well, and $v$=40 in both. (c) depicts $|\psi|^2$ of $v$=6 on a log-scale; showing that the amplitude in the inner well is extremely small. The reverse is true for $v$=7 and so on.} 
\label{fig:lifetimes}
\end{figure}

\begin{figure}
\centering
\includegraphics[clip, width=1.0\linewidth]{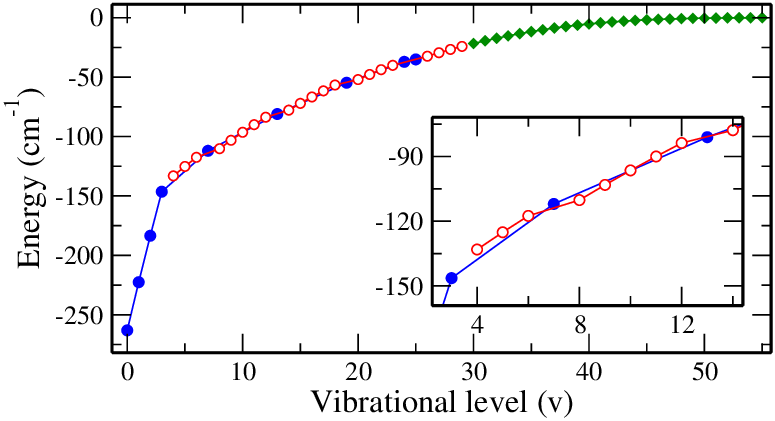}
\caption{[COLOR ONLINE] Energies of bound levels of B~$^2\Sigma_{g}^+$ state in $^{40}$Ca$_2^+$ using the same convention as in Fig.\ref{fig:lifetimes}. The inset magnifies the difference in slopes of levels localized in the inner well (shown in blue) from the ones in the outer well (shown in red).}
\label{fig:levels}
\end{figure}

We have calculated radiative lifetimes (see Fig.\ref{fig:lifetimes}) for bound vibrational levels in both the inner and outer wells of the B~$^2\Sigma_{g}^+$ state. Since these wells are separated by a large barrier, the wavefunction of the lower vibrational levels can be strongly localized in either wells. The localization of the vibrational wavefunctions can be attributed to the asymmetry of the double well (see Fig.\ref{fig:curves}) and disappears for levels above the barrier. The behavior of the radiative lifetimes exhibits $3$ distinct regimes, one where the wavefunction is mainly localized in the inner well (in blue), one where it is localized in the outer well (in red), and the last being the region (in green) in which the wavefunction spreads over both wells, resulting in poor Franck-Condon overlap with the ground X~$^2\Sigma_u^+$ state and hence longer lifetimes. The inset shows the square of the amplitude of wavefunction in $v$=$6$, of B~$^2\Sigma_{g}^+$ state in a logarithmic plot as a demonstration that the amplitude is negligible inside the inner well but still finite, preserving the correct number of nodes for that level. Fig.\ref{fig:levels} shows a plot of the energies of all bound vibrational levels in the B~$^2\Sigma_{g}^+$ state of $^{40}$Ca$_2^+$. The localization effect of wavefunctions discussed above is also exhibited in this plot; the density of levels in the more extended outer well is larger than in the inner well, leading to different energy slopes. The inset of Fig.\ref{fig:levels} exemplifies this point.

\subsection*{Electronic dipole transition moments}

\begin{figure}
\centering
\includegraphics[clip, width=1.0\linewidth]{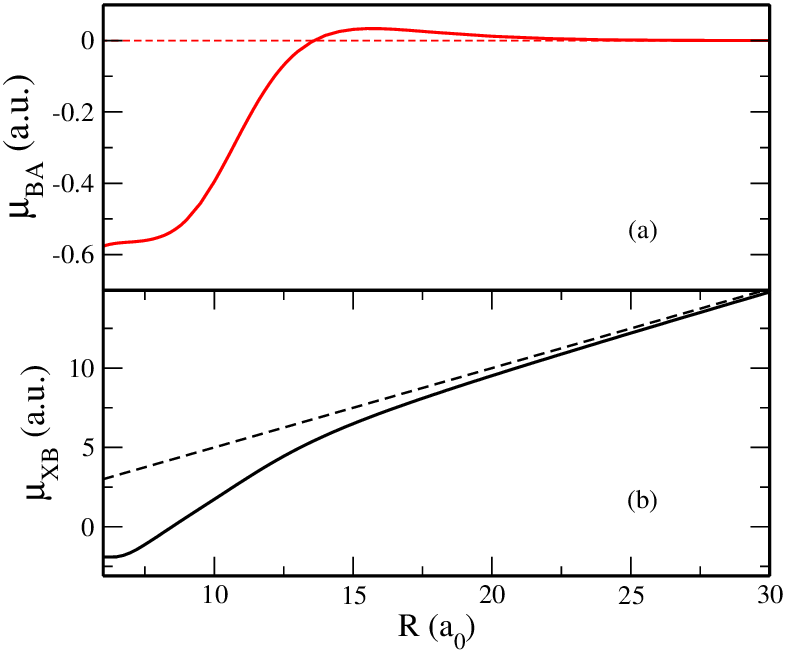}
\caption{[COLOR ONLINE] Computed electronic dipole transition moment, $\mu_{BA}$ coupling the B~$^2\Sigma_{g}^+$ to the A~$^2 \Pi_{u}$ state shown in red (a) and $\mu_{XB}$ coupling the X~$^2\Sigma_{u}^+$ to the B~$^2 \Sigma_{g}^+$ state shown in black (b). The dashed line R/2 in (b), corresponds to the classical dipole behavior.}
\label{fig:trm}
\end{figure}

For homonuclear molecules like Ca$_2^+$, there is no permanent dipole moment. However there are transitions between different electronic states which are dipole allowed. We calculate two transition moments, one of them couples the ground X~$^2\Sigma_{u}^+$ and B~$^2 \Sigma_{g}^+$ states, and the other one couples the B~$^2 \Sigma_{g}^+$ state to the excited A~$^2 \Pi_{u}$ state. To compute these transition moments, we use a complete active space self consistent field (CASSCF) wavefunction as a reference for performing multi-reference configuration interaction (MRCI) calculations. The core-valence contribution to the electronic transition moment is found to be negligible and hence omitted in the present calculations. The calculation of the transition moment coupling the B~$^2 \Sigma_{g}^+$ state to the excited A~$^2 \Pi_{u}$ state was done with the ECP+CPP valence CAS+MRCI method.\\

The electronic dipole transition moment (in atomic units) is given by,
\begin{equation}
\mu_{12} (R) = \langle 2\,|\, z\, |\, 1 \rangle \, ,
\end{equation} 

where $|1\rangle$ and $|2\rangle$ are the electronic wave functions corresponding to the pair of states X~$^2\Sigma_{u}^+$ $\leftrightarrow$ B~$^2\Sigma_{g}^+$ or B~$^2\Sigma_{g}^+$ $\leftrightarrow$ A~$^2\Pi_{u}$, when the two Ca nuclei are separated by the distance $R$. \\

Fig. \ref{fig:trm}(a) shows the electronic transition dipole moment coupling the B~$^2\Sigma_{g}^+$ and the A~$^2 \Pi_{u}$ state. The transition moment goes to zero asymptotically. Fig. \ref{fig:trm}(b) shows a plot of the computed electronic dipole transition moment between the  B~$^2\Sigma_{g}^+$ and the X~$ ^2\Sigma_{u}^+$ ground states of Ca$_{2}^+$. The transition moment $\mu_{XB}$ coupling the X~$^2\Sigma_{u}^+$ and B~$^2 \Sigma_{g}^+$ states asymptotically follows the classical dipole behavior, $\mu_{XB} \sim R/2$ \cite{Paul1977, Harvey1979}; we observe this behavior in the calculated curve of Fig. \ref{fig:trm}(b). We note that although the transition moment grows linearly with $R$, the probability of spontaneous transition will tend to zero since it is proportional to $\nu_{XB}^3$, which vanishes exponentially as $R \rightarrow \infty$.\\

\subsection*{Polarizabilities and dispersion coefficients}

\begin{table*}
\centering
\caption{The static atomic dipole, quadrupole polarizabilities and dispersion coefficient for X~$^2\Sigma_u^+$, B~$^2\Sigma_g^+$ and A~$^2\Pi_u$ states of Ca$_2^+$. All values are in atomic units. The square brackets indicate powers of ten.}
\label{tab:lr-coeff}
\begin{tabular}{cccccc}\\
\hline \hline
Molecular State   & Dipole               & Quadrupole                             & $C_4$ & Dispersion & $C_6$   \\
          & polarizability ($\alpha_d$) & polarizability ($\alpha_q$)      & (=$\alpha_d/2$) & coefficient ($c_6$)  & (=$\alpha_q/2$+$c_6$)\\ 
\hline
X~$^2\Sigma_u^+$, B~$^2\Sigma_g^+$    & $1.606 [2]$ & $3.073 [3]$  & $8.032 [1] $ & $1.081 [3]$ & 2.618[3]\\
Previous \cite{Derevianko-ca2006} 	       &  $1.571 [2]$ & $3.081 [3]$ & & & \\
Previous \cite{Mitroy} & & & & $1.085 [3]$ & \\
\hline \\
A~$^2\Pi_u$                   	             		  & $1.606 [2]$ & $3.073 [3]$ & $8.032 [1] $ & $4.950 [2]$ & $2.031 [3]$\\
Previous \cite{Mitroy2-ca2008} & &  & & \\
\hline \hline
\end{tabular}
\end{table*}

For large internuclear separations, the long-range form of the intermolecular potential can be written as

\begin{equation}
V_{LR}(R) = V_{\infty} - \sum_n \frac{C_n}{R^n} \, ,
\end{equation}

which, for the molecular ion Ca$_2^+$, can be approximated by

\begin{equation}
\label{eqn:v-lr}
V_{LR}(R) \sim V_{\infty} - \frac{C_4}{R^4} - \frac{C_{6}}{R^6} \, ,
\end{equation}

where $V_{\infty}$ is the energy of the atomic asymptote; $C_4$=$\alpha_{d}$/2, $\alpha_{d}$ is the static dipole polarizability, $C_6$ = ($\alpha_q$/2 + $c_6$), $\alpha_q$ is the quadrupole polarizability and $c_{6}$ the Van der Waals dispersion coefficient. In the expression for long range energy we have ignored the exchange energy contribution $E_{exch}$ which is very small. Also we have truncated the series at powers of $R^{-6}$, not including contributions from $R^{-8}$ and $R^{-10}$ order coefficients. \\

We have performed finite-field CCSD(T) calculations with the aug-cc-pV5Z basis set using \texttt{MOLPRO} to estimate the values of the static atomic dipole and quadrupole polarizabilities. We obtain $\alpha_{d}$ = 160.64 a.u. and $\alpha_{q}$ = 3073.39 a.u which are both in good agreement with a previous result \cite{Derevianko-ca2006} of 157.1 a.u. and 3081 a.u. respectively. The values of static dipole and quadrupole polarizability do not change for the excited A~$^2\Pi_u$ state, since it comes from an atomic asymptote in which the Ca ion is in an excited $^2D$ state whereas the Ca atom is in ground $^1S$ state (see Fig.\ref{fig:curves2}). Using a numerical fit, as described by Banerjee \textit{et al.} \cite{Banerjee2010}, we were able to extract the value of the dispersion coefficient $c_{6}$ for all the states. The value of dispersion coefficient for ground states are in good agreement with unpublished results of Mitroy \cite{Mitroy}, which are done by the methods used by Mitroy and Zhang \cite{Mitroy-ca2008, Mitroy2-ca2008} for Ca and Ca$^+$. Table [\ref{tab:lr-coeff}] lists the values of polarizabilities and dispersion coefficient for X~$^2\Sigma_u^+$, B~$^2\Sigma_g^+$ and A~$^2\Pi_u$ states of the Ca$_2^+$ dimer.

\section*{Concluding Remarks}

\textit{Ab initio} calculations have been performed on the X~$^2\Sigma_{u}^+$, B~$^2\Sigma_{g}^+$ and A~$^2\Pi_u$ states of the Ca$_{2}^+$ dimer. The calculations were computationally challenging as well as expensive because of the near degeneracy of the 4s-4p-3d orbitals in Ca. Since the B~$^2\Sigma_{g}^+ $ state has a double well, one of them being near 16.5 a$_0$, it was necessary to include diffuse functions in the basis sets to describe the well accurately. For calculating the ground X~$^2\Sigma_{u}^+$ and B~$^2\Sigma_{g}^+$ states, large augmented correlation consistent basis sets of Koput and Peterson \cite{Peterson2002} were thus chosen and the results were also corrected for basis-set superposition error. We have also corrected our valence only MRCI results for core-valence and scalar relativistic effects using CCSDT calculations with both restricted inner valence and frozen core using Peterson's cc-pwCVTZ basis set. The A~$^2\Pi_u$ state was calculated with an ECP + CPP approach similar to the previous calculations of Czuchaj \textit{et al.}\cite{Czuchaj2003}. Numerical values of the calculated potential energies are available from the authors upon request.\\

Due to lack of experimental data, we were unable to compare our theoretical values for dissociation energies or spectroscopic constants. However there are some previous theoretical results for spectroscopic constants for the X~$^2\Sigma_{u}^+$ ground state of Ca$_2^+$ \cite{Czuchaj2003, Liu1978}. These values compare well with our calculated results (see Table \ref{tab:spec}). \\

The calculation of radiative lifetimes of $^{40}$Ca$_2^+$ in the B~$^2\Sigma_{g}^+ $ state, reflects the wavefunction localization in either of the double wells. This causes the levels in the inner well to have a shorter lifetime ($\sim$ $\mu$s), whereas the ones in the outer well have much longer lifetimes ($\sim$ ms). These vibrational levels in the B~$^2\Sigma_{g}^+ $ state should generate interest for experiments in ultracold atomic and molecular physics, where these long lived molecular ions could be observed. We believe there are new prospects in both theory and experiments for atom-ion collisions, resonant charge transfer and quantum information storage using Ca$_2^+$ molecular ions.

\section*{Acknowledgements}
This work has been supported in part by the U.S. Department of Energy Office of Basic Sciences. We would like to thank Kirk Peterson for sharing his augmented functions for the correlation consistent basis sets and Jim Mitroy for sharing his calculated dispersion coefficients. We are extremely grateful to the referees for their very perceptive comments on this manuscript.\\ 

\section*{Primary Author Information}
Dr. S. Banerjee is presently a Research Engineer at Intel Corporation. His other publications can be found here 
\cite{shu2014new, smith2014experiments, banerjee2012scalable, dailey2013force, banerjee2013electronic, banerjee2013comparative, valente2013charge, shu2013emph, banerjee2012calculation, banerjee2011ab, banerjeeab, banerjee2010calculation, banerjee2009forming, banerjee2012ab, banerjee2011formation, banerjee2012qm, banerjee2011textit, banerjee2012calculation1, banerjee2010ab}.

\end{document}